\documentclass[fleqn,12pt,twoside]{article}
\usepackage{espcrc1}

\usepackage{graphicx}
\usepackage[figuresright]{rotating}

\newcommand{\AmS}{{\protect\the\textfont2
  A\kern-.1667em\lower.5ex\hbox{M}\kern-.125emS}}
  
\def \ketv #1>{\mbox{$|{#1}\rangle$}} 
\def \mate<#1|#2|#3>{\mbox{$\langle {#1}|\,{#2}\,|{#3}\rangle$}}

\title{A pentaquark model for ${\Lambda(1405)}$}

\author{T. Inoue\address{Dept. Phys. Sophia University, 7-1, Kioi-cho, Chiyoda-ku, Tokyo 102-8554, Japan}}

\begin{document}

% typeset front matter
\maketitle

\begin{abstract}
We study a pentaquark model of the negative parity hyperon $\Lambda(1405)$.
We choose a specific quark configuration for trial, 
and use the perturbative chiral quark model extended to four quarks and one anti-quark system.
We calculate a $\sigma$-term and a scalar form factor of the hyperon
and demonstrate their usefulness to study the nature of the hyperon.
\end{abstract}

\section{Introduction}

The negative parity hyperon $\Lambda(1405) S_{01}$ is established baryon resonance.
Its spin-parity, mass, and decay width are well determined experimentally:
$I(J^P) = 0(1/2^-)$, $m = 1406 \pm 4$ MeV, $\Gamma = 50.0 \pm 2.0$ MeV.
It is located at 27 MeV below $\bar K N$ threshold, and decays to $\pi \Sigma$ almost 100\%.
This hyperon is interested for many years and there are several interpretations for this hyperon.
The most simplest one is so called P-wave excited 3-quark system.
In \cite{Isgur:1978xj}, such system is studied perturbatively
with harmonic-oscillator wave function and one-gluon-exchange perturbation.
While, in \cite{Furuichi:2003eh}, the system is studied as a bound state problem.
It turns out that it is difficult to reproduce mass of $\Lambda(1405)$ in this approach.
The most popular scenario today is bound $\bar K N$ system\cite{Dalitz}
or dynamically generated resonance in meson-baryon scattering\cite{Oset:1997it}.
One can look the deep attractive potential needed to bound $\bar K N$
into $\Lambda(1405)$, for example in \cite{Akaishi:2002bg}.
In this decade, mason-baryon coupled channel scattering are studied extensively.
There, several baryons including $\Lambda(1405)$, are generated dynamically as a resonance
or a quasi bound state which decay to open channels.
Most of studies based on the chiral Lagrangian and the Bethe-Salpeter equation.
While, in \cite{Takeuchi}, the coupled channel scattering is studied at quark level,
and in \cite{Schat:1994gm}, a soliton kaon bound system is studied along the soliton picture of nucleon.
There is a lattice study\cite{Nemoto:2003ft} which shows that SU(3) flavor singlet 3-quark system
with $I(J^P) = 0(1/2^-)$, cannot be so light as $\Lambda(1405)$.

Beside these scenario, pentaquark baryon is a possible interpretation of the hyperon.
We study this scenario in this paper.
Here, pentaquark baryon stands for a state with five valence constituents, 
{\it ie.} 4-quark and 1-antiquark.
In other words, it's quark number decomposition starts form $\ketv qqqq \bar q>$ states.
By today, we do not have established pentaquark.
Indeed, $\Theta^+$ is a good candidate. But, it's existence is still in question.
If $\Lambda(1405)$ is well described as pentaquark, 
we face exotic pentaquark baryon for the first time.
In this paper, we focus on static properties of pentaquark $\Lambda(1405)$,
in stead of studying dynamical bound state problem.

The famous $\pi N$ $\sigma$-term is defined by the following formula.
\begin{equation}
%\begin{eqnarray}
\sigma_{\pi N}
  = 
      \langle N| \!\! \int \!\! d^3 \vec x ~ 
       m_u \bar u u(\vec x) + m_d \bar d d(\vec x) \, |N\rangle ~.
%  \\
% &=& \langle N| \!\! \int \!\! d^3 \vec x ~ 
%      m_u \frac{\partial {\cal H}^{\mbox{\tiny QCD}}(x)}{\partial m_u} +
%      m_d \frac{\partial {\cal H}^{\mbox{\tiny QCD}}(x)}{\partial m_d} \,
%       |N\rangle 
%\end{eqnarray}
\end{equation}
It is a measure of violation of the chiral symmetry in the real world.
Roughly speaking, it gives quark mass contribution to nucleon mass.
Similarly, $K N$ and $\eta N$ $\sigma$-terms are defined 
with the corresponding flavor combination of quark field.
Important point is that $\sigma$-terms are very sensitive to number of quark involved in baryon,
including sea-quark. 
In fact, sea-quark contribute to the $\pi N$ $\sigma$-term more than 65\% of the emprical value.
The same baryon matrix element but with finite recoil, 
is called scalar form factor.%as a function of $Q^2$.

In this paper, we study a $\sigma$-term and a scalar form factor of pentaquark $\Lambda(1405)$ particularly.
Although such data of the hyperon is not available today,
they would be useful to judge the interpretations in future.

\section{Pentaquark model}%

To make negative parity baryon with four quarks and one anti-quark,
it is simplest to put all quarks and anti-quark into the ground S-wave orbit, 
namely S-shell pentaquark.
The whole system should be color singlet for QCD quark confinement.
In addition, 4-quark subsystem should be totally antisymmetric due to the Fermi statistics. 
We can construct several flavor-spin configuration satisfying the above conditions.
In this paper, we employ one specific  flavor-spin configuration
where 4-quark subsystem is $\bf [22]$ in flavor SU(3) and $\bf [31]$ in spin SU(2).
In this combination, pentaquarks form flavor SU(3) octet and anti-decuplet.
We have a $\Lambda$ hyperon at center of octet and we study it as $\Lambda(1405)$ in this paper.
Here, spin parity of pentaquark can be $J^P = 1/2^-$ and $3/2^-$. 
Indeed, we chose former one.
In conventional 3-quark approach, $\Lambda(1405)$ is flavor singlet,
while in chiral unitary model, one pole corresponding to $\Lambda(1405)$
is almost flavor SU(3) octet\cite{Jido:2002yz}.
The present configuration is nothing but one choice.
We can study also flavor singlet pentaquark and will report it in separated paper.

We consider valence quark localized in some binding potential $S(r) - \gamma^0 V(r)$.
It's field is expanded in terms of bound states $u_{\alpha}(x)$ and $v_{\beta}(x)$.
%\begin{eqnarray}
% & & \left[i \delslash - S(r) - \gamma^0 V(r) - {\cal M} \right] \psi(x) = 0 
% \\
% & & \psi(x) = \sum\limits_\alpha b_\alpha  u_\alpha(x)
%         + \sum\limits_\beta  d_\beta^{\dagger} v_\beta(x)
%\end{eqnarray}
We set unperturbed pentaquark states with a product of the ground states,
and introduce a residual interaction by which valence quark emit/absorb
ground pseudo-scalar meson octet and perturbative gluon
\begin{equation}
 \ketv \mbox{\small Penta} >^0 = u_0(x_1) u_0(x_2) u_0(x_3) u_0(x_4) \bar v_0(x_5) ~,
\end{equation}
\begin{equation}
 {\cal L}_I = - \bar{\psi}(x) \biggl\{ S(r) i \gamma^5 \frac{\hat \Phi (x)}{F} + 
     g_s \gamma^\mu A_\mu^a(x) \frac{\lambda^a}{2} \biggr\}\psi(x) ~.
\end{equation}
Note that quark ps-meson coupling term is made so that it recover the chiral symmetry
broken by the scalar potential $S(r)$. 
We set up perturbation where baryon matrix element of a local operator is calculated by
\begin{equation}
 \mate<B|{\cal O}|B> = 
 {}^0\!\langle B| \sum_{n=0}^{} \frac{i^n}{n!}\! 
 \int \! \! d^3 \vec x \, d^4 x_1 \cdots d^4 x_n 
 T[ { {\cal L}_{I}(x_1) \cdots {\cal L}_{I}(x_n) }\, 
    {\cal O}(\vec x) \, ]
 | B \rangle^0 ~.
\end{equation}
%using the unperturbed state and the interaction.
Note that higher order terms of the quark ps-meson coupling
generate ps-meson cloud supplementing baryon.
This formalism is developed by T\"ubingen group for nucleon originally, 
and we call it perturbative chiral quark model\cite{Lyubovitskij:2000sf}.

In this model, $\sigma$-term can be studied as following.
We derive scalar density operators relevant to $\sigma$-term through Hamiltonian of the model,
%insted of QCD one, 
for example, 
\begin{equation}
  S_u^{\mbox{\tiny PCQM}}(x) 
  = \frac{ \partial {\cal H}^{\mbox{\tiny PCQM}}(x) }{\partial m_i}
  = \bar u u(x) 
      + B \left\{ \pi^+ \pi^- + \frac{\pi^0 \pi^0}{2} 
                      + K^+ K^- + \frac{\eta\eta}{6} \right\} ~.
\end{equation}
Here, first term corresponds to valence quark,
while second term does to meson cloud {\it ie.} sea-quark.
We rely on standard $\chi$SB scenario.
Then, $\sigma$-terms and scalar form factor are given by baryon matrix element
of the certain combination of these operators\cite{Inoue:2003bk}.

\begin{figure}[t]
 \includegraphics[width=0.8\textwidth]{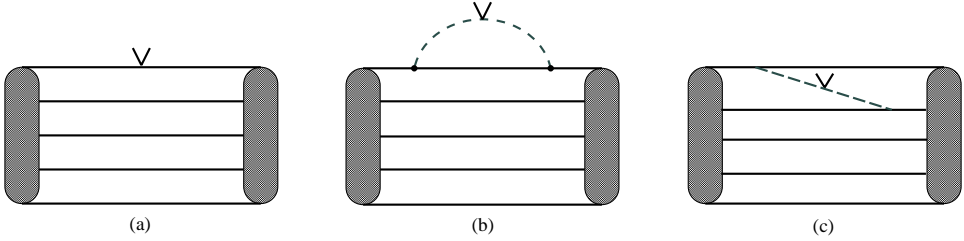}
 \caption{Perturbative series for the $\pi \Lambda$ $\sigma$-term, where $\vee$ denotes
          the density operator. }
 \label{fig:pentasig}
\end{figure}

We study the following $\sigma$-term of $\Lambda(1405)$, namely $\pi \Lambda(1405)$ $\sigma$-term.
\begin{equation}
 \sigma_{\pi \Lambda(1405)} 
  = \hat m  \langle \Lambda(1405)|
 \! \int \! \! d^3 \vec x 
 \left\{ S_u^{\mbox{\tiny PCQM}}(\vec x) + S_d^{\mbox{\tiny PCQM}}(\vec x)
 \right\} |\Lambda(1405) \rangle
\end{equation}
We evaluate the baryon matrix element perturbatively as explained above.
Fig \ref{fig:pentasig} shows the perturbative series diagrammatically.
We stop series at second order of the interaction or at meson 1-loop. 
We use quark propagator in the binding potential for internal quark line, 
but omit excited states for simplicity, and z-diagram to avoid double counting with meson.
We use standard value of quark masses $\hat m = 7$ MeV, $m_s = 25 \hat m$,
and low energy constants $B = 1.4$ GeV and $F = 88$ MeV.
For the potential binding valence quarks, 
we use a linear $S(r) = c r$ with $c=0.11$ [GeV$^2$] and a constant $V(r)$, for trial.
The resulting Dirac wave function gives proton charge radius of 0.76 [fm] at tree level.

\section{Result and discussion}

Second line of the Table \ref{tbl:sigmatb} shows $\pi N$ $\sigma$-term obtained
in the present formalism and the parameters. 
One see that empirical value of $\pi N$ $\sigma$-term is reasonably reproduced
and that pion induced sea-quark is most important.

Third line of the Table shows the $\pi \Lambda(1405)$ $\sigma$-term obtained
in the same formalism and the parameters but with pentaquark $\Lambda(1405)$.
We see that the $\sigma$-term is a little larger than the $\pi N$ $\sigma$-term.
Because number of u, d valence quark is 4 instead of 3, the contribution increase by factor 4/3. 
But, pion contribution decrease 3 MeV since number of pion loop diagram decrease a little
compared to nucleon case. 
While, number of $K$ and $\eta$ diagram increase much due to valence s quark in the hyperon, 
however their magnitudes are very small and increase the total $\sigma$-term only little.

Figure next to Table \ref{tbl:sigmatb} shows obtained $\pi \Lambda(1405)$ scalar form factor.
There, partial contributions and total scalar form factor are shown.
We see that meson cloud or sea-quark contribution disappear rapidly at finite recoil,
and form factor at $Q^2 > 0.5$ [GeV$^2$] is dominated by valence quark contribution.
This means that we can get information about number of valence quark
by seeing scalar form factor at finite recoil.
This feature is same for nucleon scalar form factor.
Therefore, if the present pentaquark picture of $\Lambda(1405)$ is realistic,
the ratio $\sigma_{\pi \Lambda(1405)}/\sigma_{\pi N}$ at medium recoil should be 4/3.

We do not have data to compare to these results.
We hope that $\sigma$-term of the hyperon will be studied experimentally
and the present results can be judged by data in future.
For this aim, we need to consider way to access $\sigma$-term of the hyperon experimentally.
Although we study only ${\pi\Lambda}$ $\sigma$-term in this paper,
${K \Lambda}$ or ${\eta \Lambda}$ $\sigma$-term could be easier to access 
and/or more useful to study nature of the hyperon.
Further studies in both theory and experimental sides
are needed to understand the interesting hyperon $\Lambda(1405)$.

\begin{table}[t]
\begin{minipage}[c]{0.50\textwidth}
\caption{Obtained $\pi N$ and $\pi \Lambda(1405)$ $\sigma$-term.}
\label{tbl:sigmatb}
\begin{tabular}{lccccc}
 \hline
 \hline
           &  val & $\pi$ &  $K$ & $\eta$ &  Total \\
 \hline
 $\sigma_{\pi N}$             &  15.4   & 34.5 & 0.95  & 0.03   & 50.9 MeV \\
 $\sigma_{\pi \Lambda}$       &  20.5   & 31.5 & 2.3   & 0.24   & 54.5 MeV \\
 \hline
 \hline
\end{tabular}
%\end{table}
%
%\begin{figure}[t]
\end{minipage}\hfill
\begin{minipage}[c]{0.48\textwidth}
%\begin{center}
 \includegraphics[width=1.0\textwidth]{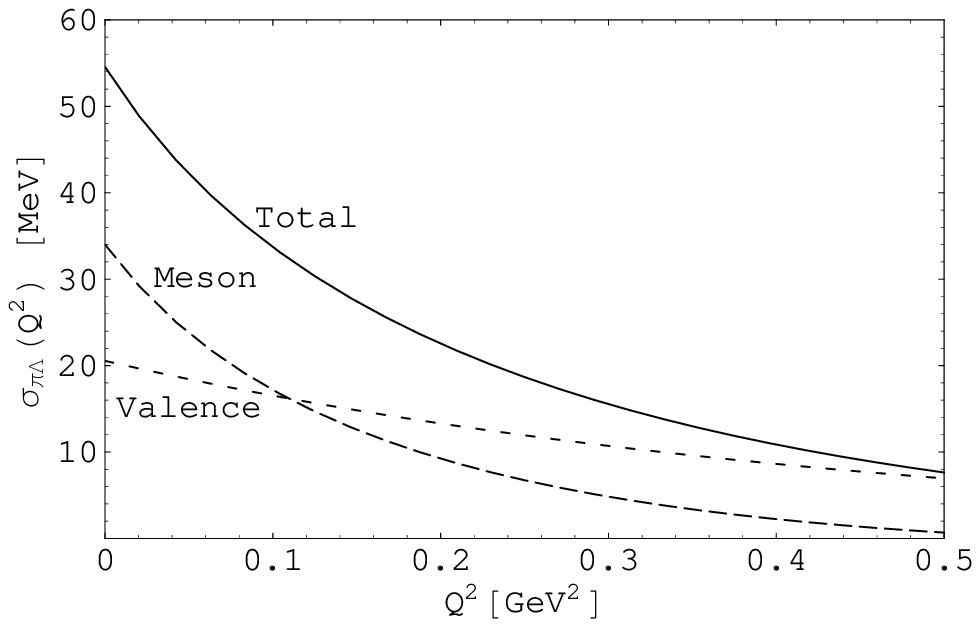}
%\end{center}
%\caption{Obtained $\sigma$-term and scalar form factor of $\Lambda(1405)$.}
% \label{fig:scffpilam}
\end{minipage}
%\end{figure}
\end{table}

% \appendix

\end{document}